\documentclass[reprint,								
               aip,										% aps layout
               jmp,										% more layout
               amssymb,									% load more AMS symbols
               amsmath,									% load AMS math (equation environment osv.)
               nobibnotes,								% if no BibTeX, then tell REVTeX normal foot
               balancelastpage,							% balance last page in two column setup
               eqsecnum, twocolumn, showkeys, superscriptaddress]								% enumeratess equations with section
               {revtex4-2}

\usepackage[latin1]{inputenc}							% letters
\usepackage[english]{babel}								% chapters etc. naming
\usepackage[T1]{fontenc}								% allow Danish inputs

\usepackage{graphicx}
\usepackage{SIunits}
\usepackage[toc,page]{appendix}
\usepackage{float}										% used to float with H
\usepackage{caption}
\usepackage[caption=false]{subfig}
\usepackage{hyperref}
\usepackage{placeins}
\usepackage[artemisia]{textgreek}
\usepackage{amssymb}
\usepackage{epsfig}
\usepackage{amsmath}
\usepackage{graphicx}
\usepackage{dcolumn}
\usepackage{latexsym}
\usepackage{color}
\usepackage{epstopdf}
\usepackage{nicefrac}
\usepackage{blindtext}
\usepackage[ulem=normalem]{changes}
\usepackage{float}
\usepackage{soul}
\usepackage{comment}
\usepackage{wedn}
\usepackage[T1]{fontenc}

\begin{document}

%\afterpage{\FloatBarrier}
\setlength{\parskip}{0pt}   %Undgaa whitespace imellem paragraffer/afsnit

\title{Collective formation of misfit dislocations at the critical thickness for equilibrium nanowire heterostructures}
%\date{\today}

\author{Thue Christian Thann}
\thanks{These authors contributed equally}
\affiliation{NNF Quantum Computing Programme, Niels Bohr Institute, University of Copenhagen, 2100 Copenhagen, Denmark}

\author{Tobias S{\ae}rkj{\ae}r$^{\dagger}$}
\thanks{These authors contributed equally}
\affiliation{NNF Quantum Computing Programme, Niels Bohr Institute, University of Copenhagen, 2100 Copenhagen, Denmark}

%\author{Toma\v{s} Stankevi\v{c}}
%\affiliation{Microsoft Quantum Materials Lab Copenhagen, 2800 Lyngby, Denmark}

\author{Sergej Schuwalow}
\affiliation{Center for Quantum Devices, Niels Bohr Institute, University of Copenhagen, 2100 Copenhagen, Denmark}

\author{Peter Krogstrup$^{\dagger}$}
%\email{krogstrup@nbi.ku.dk}
\affiliation{NNF Quantum Computing Programme, Niels Bohr Institute, University of Copenhagen, 2100 Copenhagen, Denmark}

%-------------------------------------------------

\begin{abstract}
%One of the key limiting factors for device performance in the semiconductor industry is the existence of defects associated with mismatch at heterostructure interfaces. A central parameter is the critical thickness at which an otherwise coherent thin film starts exhibiting defects associated with relaxation of the interface mismatch. 

In this work we model the evolution of strain energy during different growth stages of heterostructure nanowires. We find that the minimum energy configuration changes abruptly from fully elastically strained to partially relaxed due to collective formation of a misfit dislocation network. The transition at the critical thickness is associated with a characteristic density of misfits.
These insights are gained from a technique developed to simulate misfit dislocations in a finite element framework, incorporating both elastic and plastic relaxation in a stationary heterostructure. We argue that these results have general relevance for mismatched heterostructures.

\end{abstract}

% standard maketitle
\maketitle	

%--------------------------------------------------

\section{Introduction}

The presence of misfit dislocations (MDs) in epitaxial heterostructures alters the structural, mechanical, optical and electronic properties. The stress induced from elastic strain, originating from the mismatch between the lattice parameters of a growing thin film and the substrate, acts as a driving force for the formation of structural defects when the critical thickness is exceeded. Understanding the mechanisms that lead to formation of MDs at the critical thickness is therefore important in finding the limits of coherence for engineered epitaxial devices. The transition from elastic strain to plastic relaxation at the critical thickness in thin films has for these reasons been studied intensively in the past decades, and a number of general models for the formation of MDs have been proposed and tested\cite{MatthewsBlakeslee, Matthews, Merwe, FrankVanderMerwe1949, vanderMerwe1963-1, vanderMerwe1963-2, PeopleBean, DodsonTsao, Tsao, DodsonTsao1987, Tsao1987, Maree1987, Fitzgerald1991, Dunstan1997}. 
Most of these models examine the limit for nucleation of a singular MD in an otherwise elastically strained and pseudo-infinite planar thin film, implicitly assuming a singular MD nucleation event.\newline
%This assumption is understandable based on experimental observations, but is unproven in a theoretical and equilibrium context and will be studied here at face value
A challenge in simulating the general case of MD formation for a simple interface (pseudo-infinite system) is finding boundary conditions which reduce the model to a finite size, while not affecting the physics. Using for instance symmetric or periodic boundary conditions will hinder modeling MDs as 'additional' or 'missing' crystal planes in a pseudo-infinite film, since the outer boundaries need the ability to move freely. For this reason we choose Selective Area Growth (SAG) as a model system, given that a  full SAG nanowire (NW) geometry can be modelled to avoid challenges associated with boundary conditions for pseudo-infinite systems. Additionally, SAG offers the opportunity to design complex networks in the plane of the substrate, which makes it a promising platform for production of scale-able devices. For this reason, SAG methods for synthesis of NW heterostructure networks have received increasing interest in the field\cite{Krizek, Sole, Lee, Anna, Fahed2016, Desplanque2014, Aseev, Bakkers}.

Meanwhile, the initial growth stages of SAG NWs closely resemble the trends observed from planar films, and studies of SAG allow us to probe the mechanisms for MD formation in general. We note the important difference of a bulk section of a planar structure, which do not have have freedom to expand over the boundary and hang over e.g. a buffer structure, like SAG does. We further note rotation of crystal planes at such overhangs, as discussed further in section \ref{sec:elastic}.\newline
%In-plane SAG NW networks are constrained to underlying insulating layers, and as a consequence the crystal growth is challenged by the inherent lattice parameter mismatch, much like large planar films. 
The non-trivial morphologies found in SAG NWs may be difficult to handle in a purely analytical framework, but they are fairly easy to define with the finite element method (FEM) software employed in our model. The morphologies add an interesting study of interplay between strain relaxation along different interfacial directions, and compared to large planar films, the SAG morphology may allow the growing film an additional spatial degree of freedom, altering the critical thickness.\newline 
From studies of the free-standing Vapor-Liquid-Solid (VLS) radial type NWs, we know that critical thickness can change dramatically \cite{Ohlsson2002, Kavanagh2012, Biermanns2013} due to additional degrees of freedom from the change in morphology, with a limited NW radius in the VLS case. Theoretical models explain this phenomenon from comparison of a fully elastic model and a model exhibiting a single pair of perpendicular, interfacial MDs\cite{Ertekin2005, Glas2006}. The models predict how the critical thickness should scale with parameters such as NW radius and mismatch, and these are nicely summarized in a review by Kavanagh (2010)\cite{Kavanagh2010}. Unfortunately the VLS method lacks reliable ways of forming more complex NW networks in an experimental setting, and so our choice of SAG lets us work with a promising method somewhere between the planar thin film and the VLS NWs.\newline
Starting with the simplest case in section \ref{sec:elastic}, we study purely elastic strain relaxation in SAG NWs, to characterize the strain energy evolution for different growth stages. The shapes studied are those observed in experiments\cite{Krizek,Sole,Lee}, appearing to be approximately equilibrium shapes given lowest surface energy configuration for the NW cross section. Building on these examinations, in section \ref{section:plastic} we subsequently study elastic and plastic relaxation in SAG NWs with dislocations as 'additional' crystal planes. 
The FEM simulations are carried out for a wide range of mismatches and MD densities, finding the equilibrium configurations at the critical thicknesses from comparison between the elastic and plastic configurations. We study in particular plastic strain relaxation of a $\langle110\rangle$ orientated NW on a (001) substrate and buffer, and find a first order-like transition as a function of the extensive parameter film thickness, from a fully elastic configuration to one with a network of MDs formed collectively - a conclusion expected to carry over to other heterostructures subject to in-plane strain caused by a lattice mismatch. In later sections it will be discussed how dislocation formation observed experimentally in literature, for instance in islands, does not represent a SAG system of these conditions, and we propose an experiment to investigate our simulated predictions empirically. Lastly we analyze the stationary MD densities and show critical thicknesses and MD densities as functions of mismatch, finding a weaker dependence on mismatch than on e.g. morphology. This further allows for study of different fractions of plastic relaxation as opposed to elastic.

%--------------------------------------------------

\begin{figure*}[!ht]
%\afterpage{\FloatBarrier}
\centering
\captionsetup{width=\textwidth}

\includegraphics[width=\textwidth]{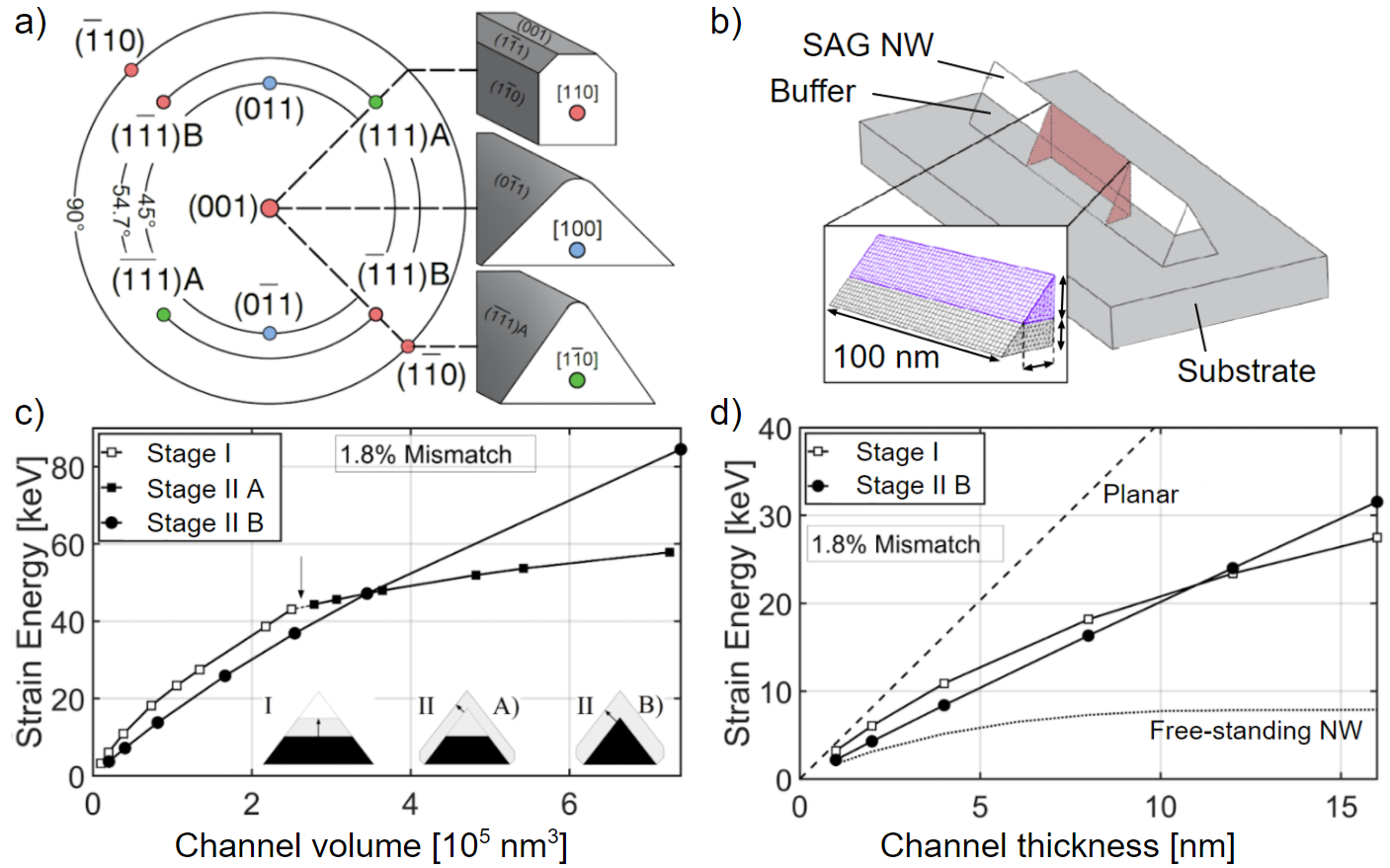}
    \caption{\textbf{Elastic growth of translationally invariant NWs.}  \textbf{a)} Linear stereographic projection of SAG NWs grown on (001) substrate. 
    \textbf{b)} Sketch of a SAG NW on a substrate with indicated symmetry planes, and a mesh example in zoom-in.
    \textbf{c)} Total elastic strain energy per 100 $\nano\meter$ section length of a $\langle100\rangle$ NW as a function of InAs transport channel volume $V_{InAs}$ on a $\mathrm{In_{0.75}Ga_{0.25}As}$ buffer (approx. 1.8\% mismatch). Insets (InAs: grey, InGaAs: black) illustrate three types of cross sectional shapes. Growth stages are described further in the main text.
    \textbf{d)} As c) except investigated as a function of the thickness of the growing layer, where the dotted line represents a free-standing NW model at same mismatch and interfacial area. The free-standing NW is simulated as hexagonal in cross-section, protruding normal to a \{111\} substrate. We consider only half a stage II B) NW, as the two sides have little to no strain field interplay.}
	\label{fig:1}
\end{figure*}

%--------------------------------------------------

\section{Purely Elastic Strain Relaxation} \label{sec:elastic}
Figure \ref{fig:1}a presents a stereographic projection of the typical NW types available on (001) substrates. The purely elastic simulation features a translationally invariant segment, using three symmetry planes as illustrated in figure \ref{fig:1}b along with an example of a preliminary mesh.
We assume for simplicity that the buffer (region separating the conducting NW channel from the substrate) is relaxed to the underlying substrate. See Supplemental Material for information on strain implementation and calculation of strain energy density (SED) in the FEM software COMSOL\cite{COMSOL}. \newline
Varying geometric parameters allows for analysis of the dependence on dimensions, shape and size of the structure. In an actual growth environment these parameters can be controlled by lithographic patterning and adjusting growth time, flux compositions and temperature. See Supplemental Material for example results of varying size effects. \newline
These simulations are run for an InAs NW on an In$_{0.75}$Ga$_{0.25}$As buffer grown in the $\langle100\rangle$ direction with $\{110\}$ side facets (see figure \ref{fig:1}a). In figure \ref{fig:1}c we show three different variations of this morphology, where stage I represents a transport channel grown from a thin layer on the buffer to a full pyramid shape with fully formed facets. We regard this shape to be an approximation of the lowest-energy shape as dictated by the surface energy densities associated with different crystallographic orientations. \newline
The other morphologies represent overgrowth, where stage II A) specifically represents a layer beginning to form on a fully grown stage I NW, and the transition is marked with an arrow in figure \ref{fig:1}c. We conclude that for our model, overgrowth contributes to total strain at a lower rate than stage I growth. We also see that stage II B) can accommodate higher mismatch for lower transport channel volumes, but also that this becomes unfavorable at larger channel volumes. All three stages exhibit sublinear increase in total strain energy for very large transport channel volumes. We note that the simulated interface area is kept constant between these morphologies for comparison. \newline
In figure \ref{fig:1}d we investigate stage I and stage II B) and the strain energy dependence on the thickness of the InAs layer. For stage II B) we consider only half of the wire (cut along the axial direction), re-dimensionalized so that the interface area is equal for both morphologies, considered a normalization to interface area. We further compare to a free-standing NW with identical interface area, which is not constrained by symmetry planes. We find this free-standing type NW to be favorable in comparison to SAG at all thicknesses, which is expected since the free-standing NW is less constrained. We note that the graphs for stage I and stage II B) cross each other at approximately 11nm in panel \ref{fig:1}d. This is due to stage I gaining less volume per unit layer thickness, as the triangular cross section becomes thinner towards the top. Hence, this crossing is absent in the panel \ref{fig:1}c displaying the energy as a function of transport channel volume. \newline
All cases compare favorably to the planar growth of thin film on a planar substrate, which is shown as the dashed line in panel \ref{fig:1}d. The thin film is a rectangular structure with symmetry planes on all four sides to emulate a pseudo-infinite plane. For the thin film case, a mismatch of 1.8\% $\mathrm{(InAs/In_{0.75}Ga_{0.25}As)}$ corresponds to a critical thickness of $\approx4.8\nano\meter$ according to Matthews model\cite{Matthews}, depending on the exact Burger's vector in the strained top layer. This highlights the morphological advantages of a SAG buffer, where the NW can relax strain by a rotational degree of freedom which has also been shown by other authors\cite{Krizek}.

%--------------------------------------------------

\begin{figure}[t]
%\afterpage{\FloatBarrier}
\centering

\includegraphics[width=\columnwidth]{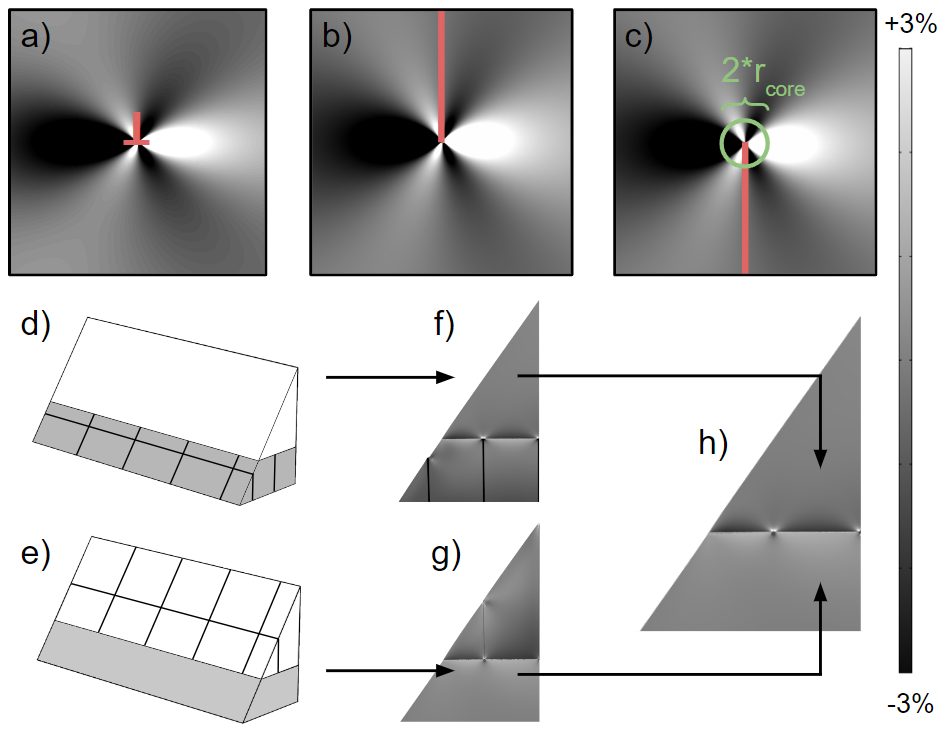}
	\caption{\textbf{FEM simulations of dislocations.} Top row: XY-components of stress fields caused by an edge dislocation at the markers. \textbf{a)} Analytical solution by Head \cite{Head}. \textbf{b)} and \textbf{c)} Results from 2D FEM simulations with dislocations modeled as a planes indicated by markers, with strain (+1) and (-1) respectively. Panel \textbf{c)} additionally shows the region near the core excluded from energy calculations, size greatly exaggerated for clarity. \textbf{d)} and \textbf{e)} 3D FEM models of $\langle110\rangle$ type NW (substrate not shown) with $\frac{a}{2}\langle110\rangle$ dislocations modeled as vertical planes with strain (+1) and (-1), respectively. \textbf{f)} and \textbf{g)} Horizontal components of strain resulting from models d) and e) with 3\% mismatch ($\mathrm{InAs/In_{0.58}Ga_{0.42}As}$ buffer). \textbf{h)} Composite image of results from the two models in panels f) and g). Colorbar applies to a-c) and f-h).}
	\label{fig:2}
\end{figure}

%--------------------------------------------------

\begin{figure*}[!ht]
%\afterpage{\FloatBarrier}
\centering
\captionsetup{width=\textwidth}

\includegraphics[width=\textwidth]{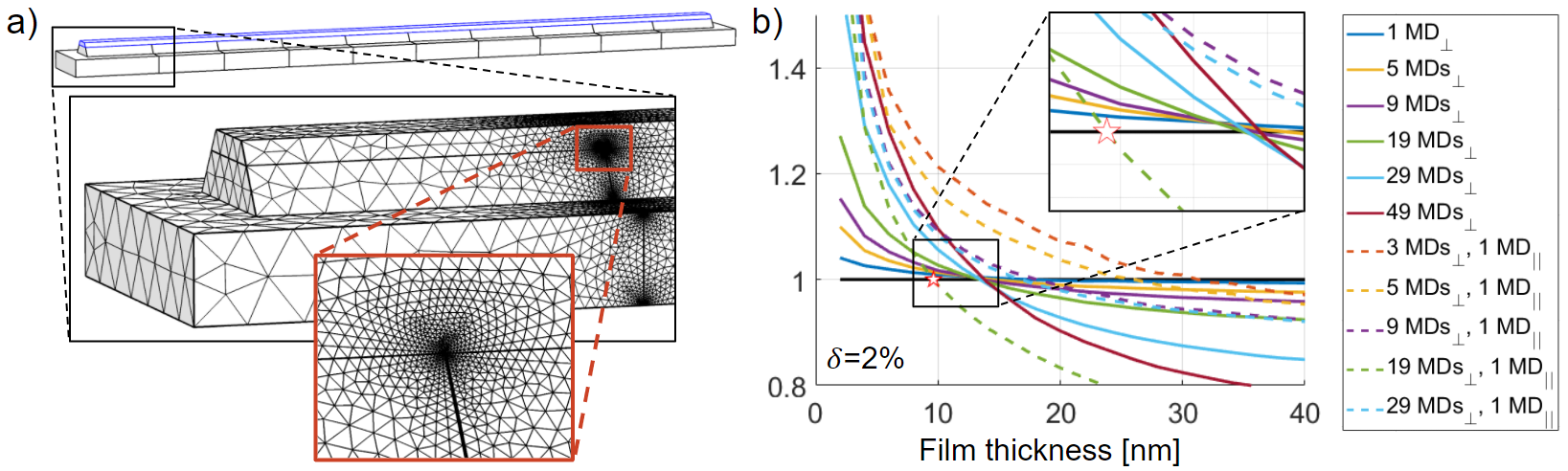}
    \caption{\textbf{Model and strain energy as function of film thickness.} \textbf{a)} Model along with a zoom section displaying a preliminary mesh with increased density near the dislocations. \textbf{b)} Total strain energy of different plastic configurations in units of the elastic configuration at the corresponding film thickness for a mismatch of $\delta=2\%$. Note that the first plastic configuration to become favorable compared to the elastic case displays a network of MDs rather than one singular MD. The dimensions of the interface in the model are $125\nano\meter$ by $4\micro\meter$.}
	\label{fig:3}
\end{figure*}

%--------------------------------------------------

\section{Plastic Strain Relaxation}
\label{section:plastic}
As the crystal volume of a lattice mismatched heterostructure increases during growth, the excess energy increases until a critical thickness is reached, at which point MDs are formed to lower the total energy, most often edge dislocations along the interfaces \cite{Krizek}. We are interested in understanding the limits of fully elastically strained heterostructures as a function of shape, volume and composition (which defines the lattice mismatch).  \newline
In a simple 1-dimensional case, the spacing between dislocations is generally given by: $d = |\vec{b}|/(\delta-\varepsilon)$, with $|\vec{b}|$ being the length of the Burger's vector, $\delta$ being the mismatch and $\varepsilon$ being the average remaining elastic strain. Therefore 'full plastic relaxation' corresponds
to $\varepsilon=0$ with a corresponding density of dislocations. However, there will be a certain fraction of elastic vs. plastic relaxation that will display the minimum strain energy, and we can not in general expect full plastic relaxation. As such, we need to examine configurations with different MD densities in order to determine the critical thickness and the associated equilibrium configuration. As a first approach the dislocations are assumed to be equidistant, but as discussed in Supplemental Information the distribution of strain may not be spatially uniform. Should material concentrations of e.g. In and Ga also be spatially non-uniform, it may be expected that dislocations form a non-equidistant network.\newline
Figure \ref{fig:2}d shows a model of a $\langle110\rangle$ type NW with edge MDs of in-plane Burger's vectors of type $\frac{a}{2}\langle110\rangle$, where $a$ is the lattice parameter. The dislocations are here modeled as planes in the buffer with thickness matching the length of the Burger's vector and positive unity strain (normal to the planes) simulating 'additional' crystal planes due to misfit dislocations at the interface. This ensures the correct effect of MDs in the transport channel, but leaves artifacts in the simulations of the buffer and substrate where the 'additional' planes should not in general be strained relative to the surrounding material far from the interface. We emphasize that this method works for finite-size structures such as the SAG morphologies chosen here, and conversely this method is incompatible with simulations of pseudo-infinite systems using fixed position symmetry planes, because they inhibit the strain-relaxing displacement generated by the additional crystal planes.\newline
An alternate method simulates the same dislocations as 'missing' planes in the transport channel with negative unity strain (figure \ref{fig:2}e). This creates the correct effect in the buffer and substrate, while the unwanted artifacts are now found in the transport channel. The dislocation planes end at the NW-buffer interface where the dislocations are situated\cite{Krizek}. In panels \ref{fig:2}a-c we compare these two methods to the analytical solution of stress fields associated with dislocations at the interface of two semi-infinite solids in 2D as found by Head (1953)\cite{Head}, which combined with mesh convergence studies give us confidence that we have reached a sufficient resolution.  While we find a clear convergence in the elastic simulations at cell sizes of 10-20\nano\meter, the characteristic cell size near the dislocation cores is on par with the length of the Burger's vector ($|\vec{b}|\approx 0.4$\nano\meter).\newline
The two methods can be combined graphically to yield the results shown in figure \ref{fig:2}h, with a complete solution being to run both simulations and consider in each only the correctly affected domains. Since by far the biggest contribution to the strain energy ($\simeq 99 \%$) is found in the transport channel on top of the buffer, we continue only with the MD model first described (figure \ref{fig:2}d) %See also Supplemental Material for further comparison between methods. \newline
\newline
As seen in figure \ref{fig:2}a-c, a small region around the dislocation cores becomes very highly strained. As a result, the elastic theory employed for evaluation of the SED is locally no longer valid, and an alternate method is needed if one wishes to evaluate the strain energy included in regions near the dislocations. We are concerned with the total strain energy in the transport channel (NW), which comprises by far the dominant energy contribution compared to the buffer and substrate. In order to evaluate the 'invalid regions' mentioned above, we modify the 'Volterra method'\cite{Merwe} or 'empirical method'\cite{HirthLothe}, excluding slightly larger cylindrical cores of radius $r_{core}=|\vec{b}|/2$, with $\vec{b}$ being the Burger's vector, arguing that the dominant energy contribution inside this range is due to the rearrangement of chemical bonds. We account for these bonds by adding an energy per unit dislocation length from the melting approximation given as $E_{m} = G b^{2}/2\pi$ where $G$ is the shear modulus of the transport channel, in this case InAs. This is likely an overestimate of the dislocation line energy density, which we will reflect upon below.
Our simulations were carried out using models of the type in figure \ref{fig:3}a (interface width $125\nano\meter$, channel length $4\micro\meter$), with 'transverse' dislocations as equidistantly spaced 'additional' planes in the substrate and then buffer. The orientation was chosen with $\langle110\rangle$ along the NW axis and \{111\} type side facets (see figure \ref{fig:1}a). The material composition of the buffer was varied with corresponding changes in material parameters according to Vegard's Law, and chosen to emulate mismatches from 1\% to 4\% corresponding to InAs on In$_{x}$Ga$_{1-x}$As with $x$ between $0.86$ and $0.44$. In all cases the composition within each region (substrate, buffer, NW) was chosen as spatially uniform for simplicity, and the distribution of dislocations as equidistant to simulate an equilibrium layer-by-layer growth as opposed to e.g. island growth. The thickness of the thin film was varied (akin to the method employed for figure \ref{fig:1}d) to emulate different stages of approximate layer-by-layer growth throughout.\newline

%--------------------------------------------------

\begin{figure*}[!ht]
%\afterpage{\FloatBarrier}
\centering
\captionsetup{width=\textwidth}

\includegraphics[width=\textwidth]{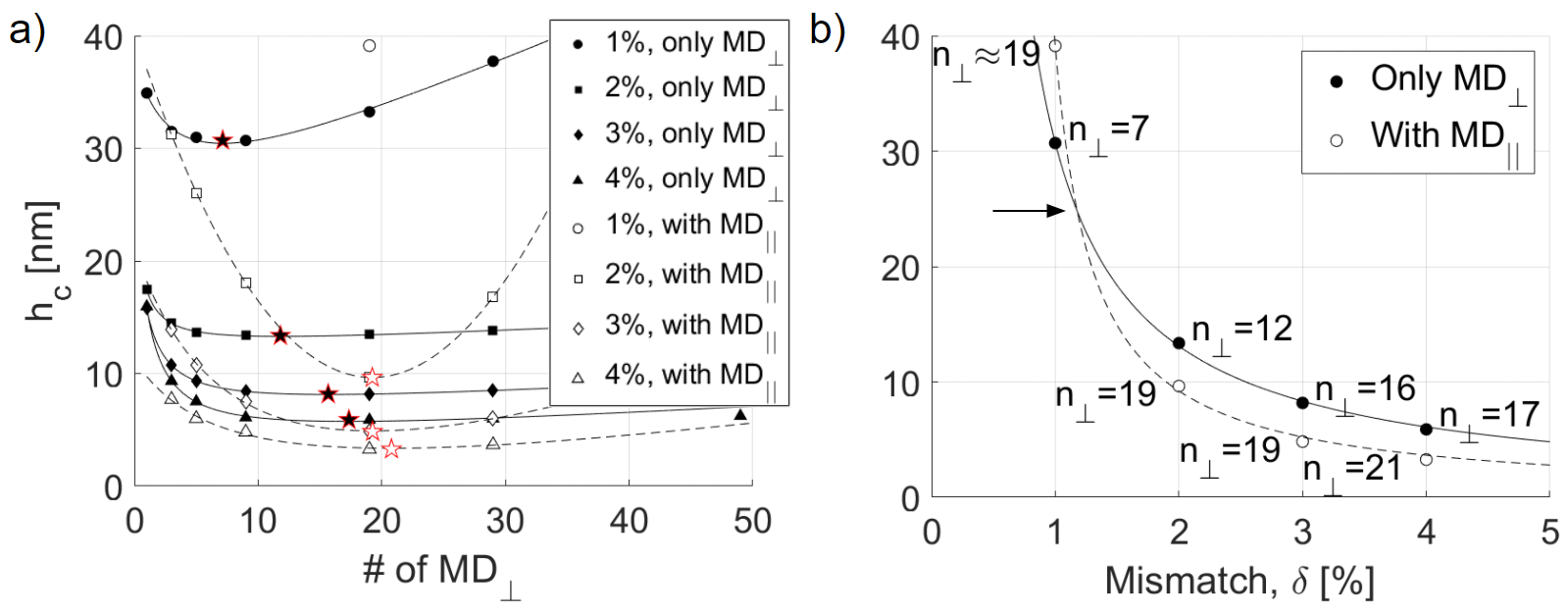}
    \caption{\textbf{Fitting critical thicknesses.} Panel \textbf{a)} shows the predicted critical thicknesses, assuming a set number of dislocations for each of the four mismatches. Markers denote minima from fits. Panel \textbf{b)} shows the minima from a) for each mismatch ($\delta$) along with the corresponding number of dislocations found from fits. Fit types are described in the main text. The dimensions of the interface in the model are $125\nano\meter$ by $4\micro\meter$.}
	\label{fig:4}
\end{figure*}

%--------------------------------------------------

\section{Results}

Figure \ref{fig:3}b shows the strain energy of plastic configurations in units of the strain energy for the purely elastic configuration as a function of film thickness for a mismatch of $\delta=2\%$. See Supplemental Material for similar results for other mismatches. \newline
From closer examination we notice the first plastic configuration to become favorable is not the one with a singular dislocation. This is a general feature across the mismatches examined, but more notable for higher mismatches. This suggests that onset of misfit dislocations at the critical thickness is a first order-like transition in an extensive parameter, to a state which becomes stable when a certain equilibrium MD density is achieved. The transition from elastically strained to a partially plastically relaxed state is characterized by both a critical thickness and a characteristic density of MDs. We note that a lower value for the dislocation line energy density would lead to configurations with a higher MD density being more energetically favored. Thus, our results for the characteristic density of MDs at the critical thickness should be considered a lower bound, based on the melting approximation mentioned in section \ref{section:plastic} above.\newline
In this study we have limited ourselves to one axial dislocation running along the center of the NW, and note that a more complete examination would have to deal with a much larger parameter space of both number and positions of axial MDs.\newline
We also note the general feature that higher mismatches tend to favor configurations with more MDs. For the 1\% mismatch case the equilibrium configuration at the critical thickness has only transverse dislocations (MD$_{\perp}$), which could be interesting for engineering of MDs in heterostructures. However, for the mismatches of 2\%, 3\% and 4\% the equilibrium configurations at the critical thicknesses have both the axial and transverse dislocations present. This could prove useful for analysis, as the lack of an axial MD from cross sectional TEM of a high mismatch structure could indicate that the entire structure is purely elastically relaxed.\newline
The question of MD configuration at the critical thickness is examined further in figure \ref{fig:4}a which shows the points where different configurations become favorable compared to the purely elastic case. For a given mismatch the lowest of the critical thicknesses is the predicted equilibrium critical thickness, and a specific MD density is associated with this.\newline
The guidelines in figure \ref{fig:4}a are fits to the form
$h_{c} = an_{disl} + b + c/(n_{disl}+d)$, where $n_{disl}$ is the number of dislocations. The minima from figure \ref{fig:4}a (marked) are extracted from the fits and plotted in figure \ref{fig:4}b along with the associated number of MDs and new fits of the simpler form $h_{c}=\alpha/(\delta+\beta)$ where $\delta$ is again the mismatch. \newline
The variables $\alpha_{\perp}=22.8\nano\meter$, $\beta_{\perp}=-0.26\%$, $\alpha_{\mid\mid}=12.1\nano\meter$, and $\beta_{\mid\mid}=-0.69\%$ are found from the fits for configurations without and with the axial dislocation, respectively. For mismatches below $\delta_{\mid\mid}^{*} = 1.2\%$ (marked by an arrow in figure \ref{fig:4}b), the configuration at the critical thickness shows no axial dislocation. Interestingly, the density of transverse MDs at the critical thickness only increases slightly while increasing the lattice mismatch from 1\% to 4\%. In the entire range, spacing between MDs is found to be around 200\nano\meter, corresponding to a partial plastic relaxation of approximately 0.2\% misfit strain (emphatically not 0.2\% \textit{of} misfit strain). In the case of 2\% misfit we thus find the fraction of plastic relaxation to be 1/10. The weak dependence on misfit strain and low value suggest the equilibrium configuration at the critical thickness is more dependent on e.g. morphology, and the melting approximation overestimating the dislocation line energy density. Further work for investigating this is discussed below.\newline
The fit forms and variables found suggest a divergence of the critical thickness at a mismatch of a quarter of a percent. To ensure a fully elastic growth in stage 1 however, it is only necessary for the critical thickness to be larger than the thickness of the transport channel grown. Due to the geometry chosen for the model, the stage 1 transport channel can grow to a maximum thickness of $h_{max}=w/\sqrt{2}$, where $w=125\nano\meter$ is the width of the interface. In our model this can be accommodated elastically at a mismatch of $\delta^{*} = 0.52\%$, meaning a buffer of In$_{0.93}$Ga$_{0.07}$As. 
While a buffer this high in In concentration may cause issues in containing the wavefunction to an intended transport channel, we note the height of $125\nano\meter$ is only an example, as is the chosen elements of In, As and Ga. We note that while all the critical thicknesses quoted are specific to the morphology, dimensions and materials, the method presented can be used for examination of other combinations and structures.

%--------------------------------------------------

\section{Discussion and Conclusion}

We find the mechanisms of strain relaxation in lattice-mismatched SAG NWs to be distinctly different from reports in literature on planar heterostructures and on free-standing NWs\cite{Ertekin2005, Glas2006}.
Compared to planar thin films, the additional elastic relaxation for SAG stems from the rotational degree of freedom for relaxation transverse to the NW axis which in principle can overshoot the bulk relaxed values, giving additional room for elastic relaxation along the NW axis through the Poisson effect. We identify three different growth stages, all of which are energetically favorable compared to planar thin film growth, and all of which are sublinear but quickly become approximately linear with different dependencies on layer thickness, favoring stage II A). \newline
Our findings establish a relationship between transport channel layer thickness and MD density for a SAG NW morphology, similar to that between NW radius and misfit percentage as found by Ertekin \textit{et al.} (2005)\cite{Ertekin2005} and Glas (2006)\cite{Glas2006} for VLS NWs. This highlights the difference between SAG and free-standing NWs. For comparison we quote the experimentally found critical thickness of $h_{c,film\%}=1.71\nano\meter$ for planar thin film growth of InP on GaAs at 3.8\% lattice mismatch \cite{Mazuelas}. This shows the ability for elastic relaxation in SAG NWs as somewhere between the highly constrained planar thin films and the nearly unconstrained free-standing NWs of VLS. \newline
We compare to previous efforts in using FEM to analyze misfit dislocations such as Ye \textit{et al.} (2009)\cite{Ye2009}, which also use initial strain as a numeric technique, but fails to include both elastic and plastic relaxation simultaneously and naturally does not take the spatial freedoms and lattice directions of SAG NWs into account. Therefore we believe our methods are novel and relevant for finite-size morphologies across materials, and the results can be compared to physical samples by analysis of e.g. atomic resolution TEM with GPA\cite{Liu2013, Peters2015}.\newline
It is appropriate to discuss this abrupt and collective formation in relation to observed formations. As done by LeGoues et al (1994)\cite{cyclic}, UHV TEM can be used \textit{in situ} to observe singular MD formations in island growths, with accompanying changes in growth velocity immediately before and after nucleation events. 
They provide rudimentary theoretical considerations with equilibrium assumptions, but we consider their model incomplete as it does not include strain energy as a driving force and does not explain the preference for island growth over layer-by-layer.
We understand that in our study we have specifically simulated an equilibrium environment, and created a rigorous framework incorporating strain energy and additional references. \newline
A later study by Merdzhanova et al (2006)\cite{dendro} uses a more time-efficient AFM method, and notes higher growth temperatures consistently giving rise to more singular nucleation events, as well as a dramatic change in the size of islands. Particularly the balance between coalescence of neighbouring islands growing simultaneously, as opposed to islands growing smaller when located in the depletion zone of a larger island, is affected.  They do not present a complete theory for this behavior, but suggest a scenario qualitatively involving material intermixing. \newline
We note these points to be different from the assumptions within our model, and that temperature dependencies agree well with a non-equilibrium nature of the process. As such we find the discrepancies to our model as expected, and they underline the potential in understanding the equilibrium and non-equilibrium divide in phenomena and behavior. In particular, the role of material intermixing could be introduced and studied within our framework, posing an immediate candidate for further work. Among other things it would affect the spatial distribution of strain, which may cause the optimal network of dislocation to be non-equidistant. \newline
Additionally it would be possible to design an experiment using low growth rate, high temperatures and low mismatch, which would be a better representation of the simulation in reality. High energy presumably allows for breaking of kinetic barriers and avoiding local minima, approaching our predicted global minimum of collective formation, and low mismatch with a low growth rate allows for true layer-by-layer growth, seeing as the island growth seen in literature cannot represent this system. We propsoe using in-situ observation of strain in the layer-by-layer growth, for instance using a curvature tool as done by Gilardi et al (2018)\cite{Gilardi2019}, to investigate to what degree the stationary (non-time dependent) assumptions of the model are correct, when the other conditions as described above are fulfilled. We expect the results to be strictly different from the discussed literature, where the conditions are not fulfilled.
\newline

In summary we present a novel method for introducing plastic relaxation from MDs as localized FEM features in heterostructure simulations, allowing an examination covering different morphologies and MD densities. This leads to our prediction of collective rather than singular onset of MDs at the critical thickness, which is a novelty. For SAG NW growth in stage I, we find critical thicknesses of $h_{c,1\%,\perp} = 30.7\nano\meter$, $h_{c,2\%,\mid\mid} = 9.6\nano\meter$, $h_{c,3\%,\mid\mid} = 4.8\nano\meter$, and $h_{c,4\%,\mid\mid} = 3.3\nano\meter$ for 1\%, 2\%, 3\%, and 4\% mismatch, respectively, as summarized in figure \ref{fig:4}. In all cases we find that collective formation, as a first order-like transition as a function of film thickness, is favorable compared to singular onset. For mismatches below $\delta^{*}_{\mid\mid} = 1.2\%$ we find that the equilibrium configuration shows only transverse dislocations, while for mismatches above this value both axial and transverse dislocations are expected. At the critical thickness, the density of MDs suggests initial plastic relaxation of approximately 0.2\% misfit strain in the range of misfits examined. Further studies are needed in order to examine in more detail how this initial plastic relaxation changes with morphology, compositions and different values for the dislocation line energy density. 
We argue that our results are relevant for general heterostructures, predicting that a first order-like transition in our finite-size case carries over to e.g. a pseudo-infinite planar heterostructure. %As discussed in \ref{section:plastic} there is agreement with the analytical solution for semi-infinite planes, while it is not possible to use fixed position symmetry planes, because they create unphysical effects when introducing misfit dislocations.

%While the simulation method is incompatible with fixed position symmetry planes, our study of finite-size heterostructures predict a first order-like transition in the extensive film thickness parameter in general for mismatched heterostructures.\newline

%As discussed in section \ref{section:plastic}, this simulation does not utilize fixed position symmetry planes typically used as boundary conditions in simulations of pseudo-infinite planar heterostructures, because they create unphysical effects when introducing misfit dislocations. We argue that our results remain relevant for general heterostructures, predicting a first order-like transition in the extensive parameter of film thickness for finite-size structures carries over to e.g

%--------------------------------------------------

\section*{Acknowledgements}

This paper was supported by the European Research Council (ERC) under grant agreement No. 716655 (HEMs-DAM). We also thank the entire Center for Quantum Devices at University of Copenhagen, which housed us for a time and served as the original affiliation until most authors moved on to NQCP.

We also enjoyed practical support from Microsoft Station Q through their connection to Qdev, especially the guidance of Toma\v{s} Stankevi\v{c} in the use of COMSOL. We thank him for many discussions and his wealth of knowledge.

The authors also thank Mart\'{\i}n Espi\~{n}eira, Daria Beznasyuk, Anna Wulff Christensen, Filip K\v{r}\'{\i}\v{z}ek, Thomas Kanne, Joachim Sestoft, Jordi Arbiol, Sara Mart\'{\i}-S\'{a}nchez, and Philippe Caroff for scientific and practical inputs.\newline

%Kevin van Hoogdalem
%L\'{e}o Bourdet

%--------------------------------------------------

% custom contributions and email

\noindent %$^{*}$: These two authors contributed equally. \newline
$^{\dagger}$: tobias.saerkjaer@nbi.ku.dk, krogstrup@nbi.ku.dk

%--------------------------------------------------

\end{document}

%\preprint{APS/123-QED}

\title{Supplemental Information: Collective formation of misfit dislocations at the critical thickness for equilibrium nanowire heterostructures}% Force line breaks with \\
\date{\today}% It is always \today, today,

\author{Tobias S{\ae}rkj{\ae}r$^{\dagger}$}
\thanks{These authors contributed equally}
\affiliation{NNF Quantum Computing Programme, Niels Bohr Institute, University of Copenhagen, 2100 Copenhagen, Denmark}

\author{Thue Christian Thann}
\thanks{These authors contributed equally}
\affiliation{NNF Quantum Computing Programme, Niels Bohr Institute, University of Copenhagen, 2100 Copenhagen, Denmark}

\author{Sergej Schuwalow}
\affiliation{Center for Quantum Devices, Niels Bohr Institute, University of Copenhagen, 2100 Copenhagen, Denmark}

\author{Peter Krogstrup$^{\dagger}$}
\email{krogstrup@nbi.ku.dk}
\affiliation{NNF Quantum Computing Programme, Niels Bohr Institute, University of Copenhagen, 2100 Copenhagen, Denmark}

%--------------------------------------------------

\begin{comment}

    \begin{abstract}
    An article usually includes an abstract, a concise summary of the work
    covered at length in the main body of the article. 
    
        \begin{description}
        \item[Usage]
        Secondary publications and information retrieval purposes.
        \item[PACS numbers]
        May be entered using the \verb+\pacs{#1}+ command.
        \item[Structure]
        You may use the \texttt{description} environment to structure your abstract;
        use the optional argument of the \verb+\item+ command to give the category of each item. 
        \end{description}
    
    \end{abstract}

%--------------------------------------------------

\pacs{Valid PACS appear here}% PACS, the Physics and Astronomy
                              
\end{comment}

\maketitle

%--------------------------------------------------

\section{COMSOL models for FEM Simulation}
Our FEM simulations were carried out as 3-dimensional Stationary studies with the Solid Mechanics part of the Structural Mechanics module in COMSOL Multiphysics\cite{COMSOL}.
Materials in COMSOL can be defined with a variety of different properties, either from scratch or from a library of predefined materials. The important properties are the Bulk Modulus, Poisson ratio and elasticity matrix. The entries of the elasticity matrix are used for calculations of strain energy density from linear elastic theory.
The Linear Elastic Material sub-menu of the Solid Mechanics part of the interface allows for imposing initial strain in select parts of any geometry built.

Initial tensile strain is employed (in-plane with the interface) in the NW corresponding to a chosen lattice mismatch between NW and the buffer, from which a balance of forces on each mesh point yields the final configuration with forced coherence at the interface. 

The strain energy density (SED) is found locally from derivatives of the displacement according to equation \ref{eq:SED}:

\begin{equation}
    U_{SED} = \sum\limits_{ijkl} \frac{1}{2} c_{ijkl} \varepsilon_{ij} \varepsilon_{kl}
    \label{eq:SED}
\end{equation}

with $c_{ijkl}$ being the stiffness coefficients and $\varepsilon_{ij}$ components of the strain tensor. Evaluations of the resulting strain must properly account for the initial strain imposed. \newline
The relevant bulk parameters are the lattice and elastic constants and for In$_{x}$Ga$_{1-x}$As, where we assume linear interpolation between the respective parameters\cite{LatInAs, LatGaAs} of the component materials (Vegard's law). \newline

Some drawbacks do arise from the static and continuous simulations of a dynamic and atomistic physical system, and we should address those here. Drawbacks include but are not limited to a lack of polarity, static elasticity and lattice constants, and uniform compositions within each region. While these drawbacks are relevant and present, plenty of results are still obtainable, and simulations of strain relaxation and dislocations in SAG heterostructures could prove a central tool for achieving dislocation free, scalable, high mobility devices.

%--------------------------------------------------

\begin{comment}
\section{Geometric sweeps}

The COMSOL framework allows us to easily sweep over any range of values for any parameter. It may be of particular interest to see how the strain energy integrated over a pre-defined bulk section, e.g. a cube of sidelength $3\nano\meter$, in the center changes as a function of geometric parameters, such as the height of the buffer and the length of the entire heterostructure. This allows for investigation of the energy in different sectors of the heterestructure, and at which dimensions that value converges. Such runs for a $[100]$ type NW at 3\% mismatch is seen in Figure \ref{fig:geometricsweeps}.
This method can be employed for examination of parameters in both dislocated and non-dislocated configurations.

\begin{figure}[h!]
\centering
%\afterpage{\FloatBarrier}
\includegraphics[width=\columnwidth]{GeoSweeps.jpg}
	\caption{The total strain energy in a pre-defined cubic section in the center of the heterostructure, as a function of buffer height and length of the wire. Here it has been run for a $[100]$ NW and we see flattening of the curves, suggesting that size effects are negligible beyond.}
	\label{fig:geometricsweeps}
\end{figure}

\end{comment}

\section{Finite Length Effects}
\label{section:end}
\begin{figure}[h!]
%\afterpage{\FloatBarrier}
\centering
\captionsetup{width=0.8\columnwidth}
\includegraphics[width=0.8\columnwidth]{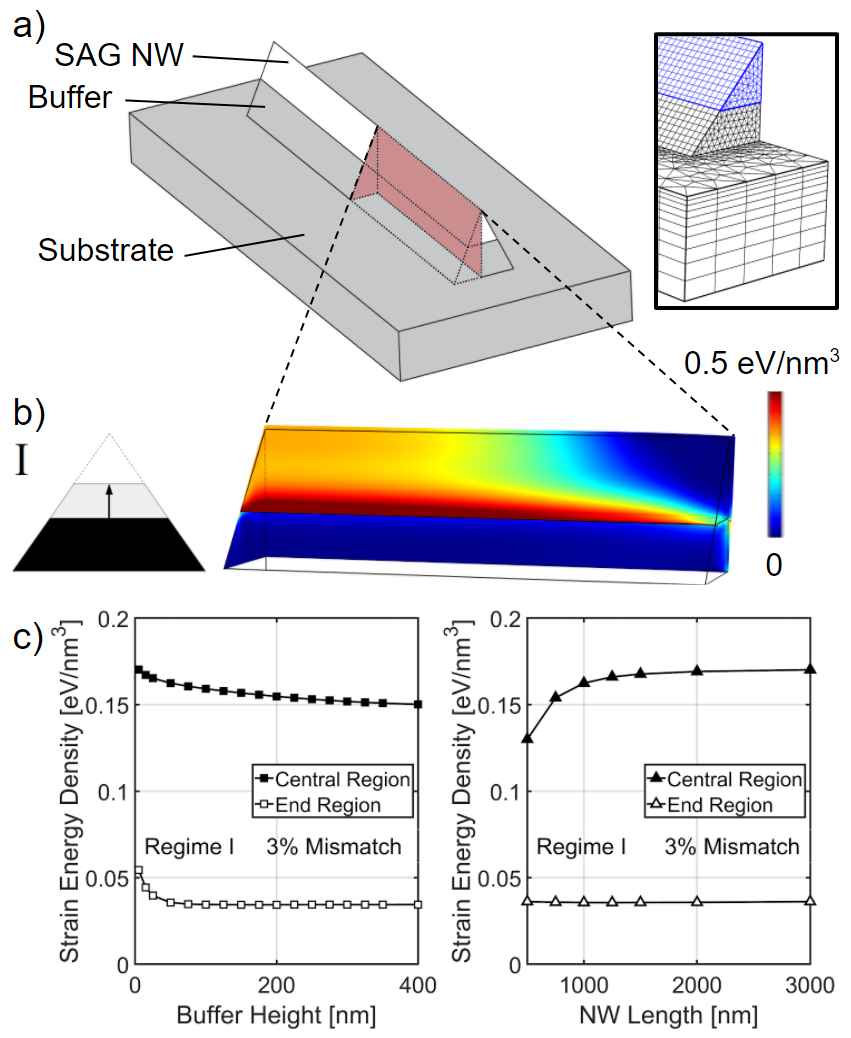}
	\caption{\textbf{Finite length effects, 3\% mismatch $\mathrm{(InAs/In_{0.58}Ga_{0.42}As)}$.} \textbf{a)} Illustration of end section with symmetry planes. \textbf{Box:} Example mesh of end section. \textbf{b)} Distribution of strain energy density on buffer-NW interface as well as both symmetry planes for a fully triangular shaped NW (fully grown stage I). \textbf{c)} Strain energy in a central cube as a function of size parameters, showing length scales for decoupling of the center to the end effects.}
	\label{fig:s1}
\end{figure}

Turning our attention to the regions near the NW ends, figure \ref{fig:s1}a shows a NW morphology for stage I growth with only two symmetry planes imposed to examine finite length and the corresponding gradient in elastic strain energy density towards the end of the NW. \newline
The translationally invariant NW segments investigated in the main text relax strain primarily by rotation of crystal planes in directions perpendicular to the NW axis. Near the ends of the NWs, rotation along the NW axis provides an additional degree of freedom for relaxation. \newline
Figure \ref{fig:s1}b shows the distribution of SED in the stage I model. Unsurprisingly, the general trend shows a higher SED near the interface tapering off with distance. For the stage I growth we see the SED decreasing as we move from the middle towards the NW sides. \newline 
Figure \ref{fig:s1}c shows a comparison of the average strain energy density of $50$nm sections in the center and end of the NW. The simulations are run for a $\langle100\rangle$ type NW at 3\% mismatch $\mathrm{(InAs/In_{0.58}Ga_{0.42}As)}$ NW, and we investigate the geometric parameters of NW length and buffer height.\newline
We notice a clear trend of the end region converging faster and at lower values. We also see target dimensions of a buffer in order to minimize strain energy. This method can be employed for examination of parameters in both fully elastic and plastic configurations.

%--------------------------------------------------

%--------------------------------------------------

%--------------------------------------------------

\newpage

\section{Strain Framework}

For pedagogical reasons, the main paper and this supplemental is usually phrased in terms of linear strain, that is:
\begin{equation}
    \varepsilon_{l} = \frac{L}{L_{0}} - 1
    \label{eq:linstrain}
\end{equation}

with $\varepsilon_{l}$ being the linear strain and $L$ being the final length of an object with unstrained length $L_{0}$. In this linear strain framework, the strain is just the fractional elongation of the object. The reader should note, that several frameworks for strain are available, and notably that strain in the Solid Mechanics module of COMSOL\cite{COMSOL} is Green-Lagrange strain:

\begin{equation}
    \varepsilon_{GL} = \frac{1}{2} \left[ \left( \frac{L}{L_{0}} \right) ^{2}-1 \right]
    \label{eq:glstrain}
\end{equation}

The reason is that GL-strain is appropriate for larger strains where linear approximations do not function. Other FEM packages may employ different strain frameworks such as Almansi or logarithmic strain, and adequate adjustments should be taken to account for this. E.g. we note that $\varepsilon_{l}=+1$ corresponds to $\varepsilon_{GL}=3/2$, while $\varepsilon_{l}=-1$ corresponds to $\varepsilon_{GL}=-1/2$. For the purposes of this supplemental, we shall continue to phrase strain in the linear terms, since it allows for a more intuitive understanding, keeping in mind that the specific implementation is recast depending on the strain framework of the FEM software.
\begin{figure}[h]
\centering
\includegraphics[width=0.95\columnwidth]{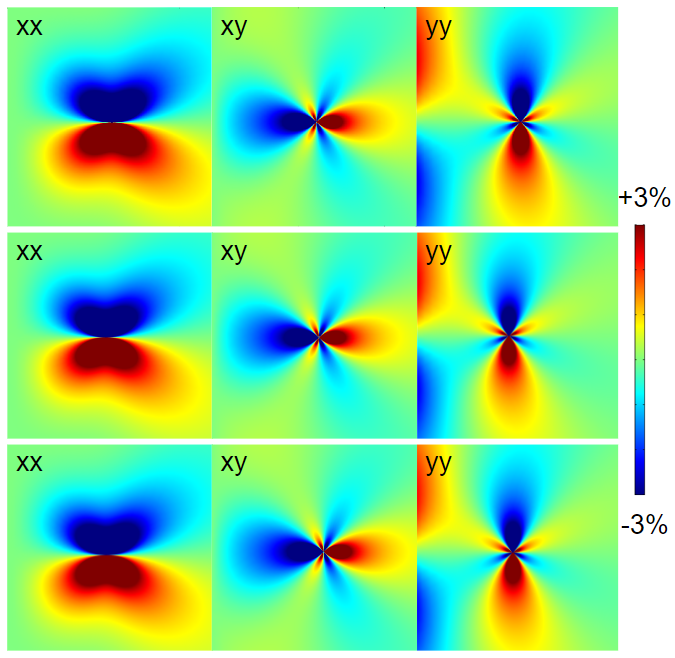}
	\caption{Stress components caused by an edge dislocation near a fixed surface (left side). Top row: Analytic solution by Head \cite{Head}. Mid row: Simulation using $(+1)$ strain method. Bottom row: Simulation using $(-1)$ strain method. The middle column is also shown in the main text.}
	\label{fig:s2}
\end{figure}

%--------------------------------------------------

\section{Analytic Solution and Method Equivalence}
The default simulation including MDs modeled as planes with $+1$ strain in the buffer, corresponding to the 'additional' crystal planes. As mentioned in the main text, this ensures correct boundary conditions from the interface and above, while the boundary conditions inside the lower part are obviously incorrect, since the imposed planes are not actually strained compared to the surrounding material, as confirmed in e.g. geometric phase analysis of atomic resolution TEM. This method is especially viable, since we are mostly concerned with the variations in the transport channel.
Equivalently we can model the MDs as planes with $-1$ strain in the wire, corresponding to the 'missing' crystal planes. This method ensures correct boundary conditions from the interface and below. \newline

\begin{figure}[h]
\centering
\includegraphics[width=0.95\columnwidth]{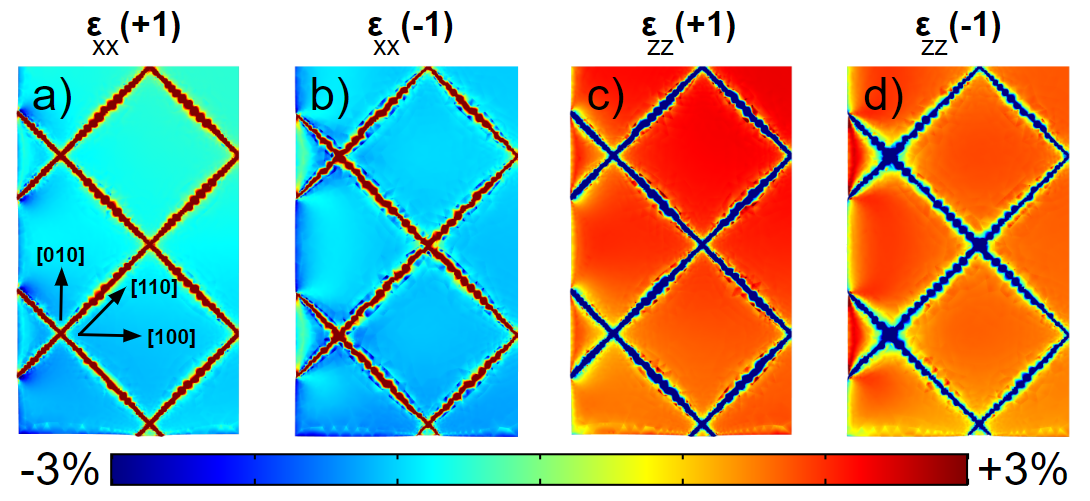}
	\caption{Top views of a section of the NW at the interface between NW and buffer for a \{100\} type NW. a) Transverse (horizontal) strain component using the method with $+1$ strain in the buffer. b) Same as a, using the method with $-1$ strain in the NW. c) Out of plane strain component, $+1$ strain method. d) Same as c, using $-1$ strain method.}
	\label{fig:s3}
\end{figure}

\begin{figure*}[h!]
\centering
\captionsetup{width=\textwidth}

\includegraphics[width=0.95\columnwidth]{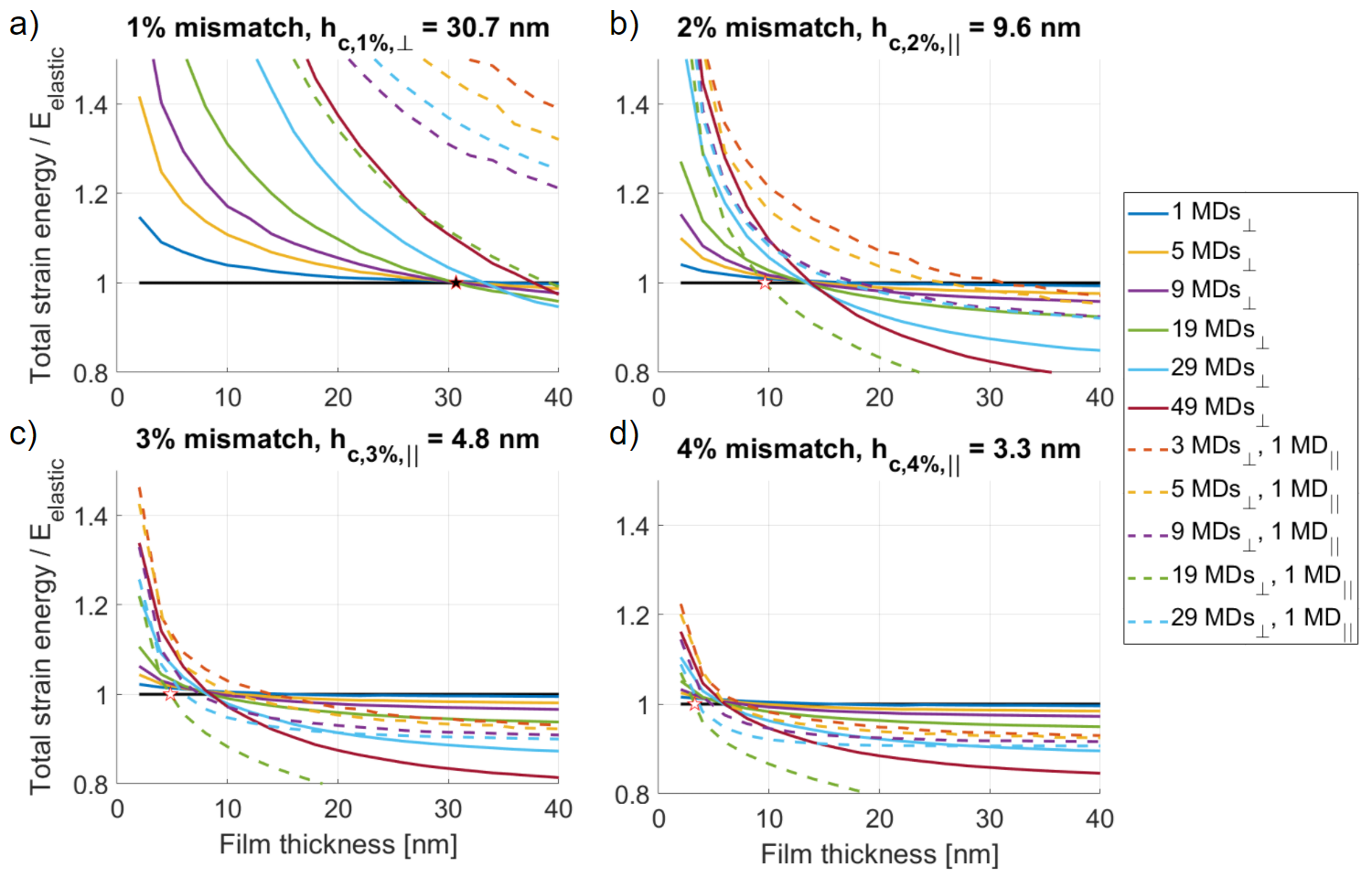}
    \caption{\textbf{Strain energy as a function of film thickness.} Each panel shows an examination for a specific mismatch. For each film thickness the favored configuration is the one with the lowest strain energy, and the critical height is found when a plastic configuration crosses below the unity line (first is marked).}
	\label{fig:s4}
\end{figure*}

The simulation results should agree with analytic solutions, e.g. solutions by Head \cite{Head}. Fig. \ref{fig:s2} shows different components of a 2D stress fields caused by an edge dislocation, as analytic solution or modeled by the methods mentioned above.
The figure shows a clear equivalence between all three methods, validating the simulational method. For the purposes of our simulation the concept is straight-forwardly extended to three dimensions.
Dislocations are associated with an energy, which is proportional to the square of the length of the Burger's vector. This makes the lowest energy MDs of zinc-blende those with Burger's vectors of type $\frac{a}{2}\langle110\rangle$. For a $\langle 100 \rangle$ NW on a (001) substrate, these types of MDs will simultaneously relax strain in two out of the three directions: axial, transverse and out of plane. In the case of a $\langle 110 \rangle$ NW on a (001) substrate, there exist favorable MDs which relax strain in distinctly axial, transverse or out of plane directions as well as MDs which relax all directions simultaneously. This was illustrated in the main text with the model for formation of MDs. The method proposed here is equally well suited for MDs with other directions of Burger's vectors, as the strain associated with the MD can be defined independently of the MD plane.
%--------------------------------------------------

\section{Composite Plots}
Combining the two methods mentioned above (strain $+1$ and $-1$) as one composite solution relies on the two solutions producing a consistent solution at the interface. Figure \ref{fig:s3} shows resulting strain components at the interface of a simulation of a $\langle 100 \rangle$ type NW on a (001) substrate, where the initial boundary conditions are considered equivalent. The $\langle 110 \rangle$ type NW as well as other strain components were examined as well with similar agreement between methods.
%--------------------------------------------------
\section{Energy Results for Plastic Configurations}
The results of the strain energy integrations for the plastic configurations with the axial MD (MD$_{\mid\mid}$) are shown in figure \ref{fig:s4}. As is evident, the equilibrium configuration at the critical height shows no axial MD (MD$_{\mid\mid}$) for the 1\% mismatch (a solid line crosses below the unity line first), while for 2\%, 3\% and 4\% the equilibrium configuration at the critical height shows at least one axial MD (a dashed line crosses is the first to cross below the unity line). The results are summarized in the main text. 
\newline
\newline
\newline

%--------------------------------------------------

\noindent %$^{*}$: These two authors contributed equally. \newline
$^{\dagger}$: tobias.saerkjaer@nbi.ku.dk, krogstrup@nbi.ku.dk

%--------------------------------------------------

%\cleardoublepage